\documentclass{article}
\usepackage[T1]{fontenc} 
\usepackage[utf8]{inputenc} 
\usepackage{ismir,amsmath,cite,url}
\usepackage{graphicx}
\usepackage{color}

\title{MIDI--Sheet Music Alignment Using Bootleg Score Synthesis}

\multauthor
{Thitaree Tanprasert$^{1*}$ \hspace{1cm} Teerapat Jenrungrot$^{1*}$ \thanks{${^*}$ The first two authors had equal contribution.} \hspace{1cm} Meinard M{\"u}ller$^2$ \hspace{1cm} TJ Tsai$^1$} { 
  $^1$ Harvey Mudd College, Claremont, CA USA\\
$^2$ International Audio Laboratories Erlangen, Germany\\
{\tt\small \{ttanprasert, mjenrungrot, ttsai\}@hmc.edu, meinard.mueller@audiolabs-erlangen.de}
}
\def\authorname{Thitaree Tanprasert, Teerapat Jenrungrot, Meinard M{\"u}ller, TJ Tsai}

\begin{document}

\maketitle
\begin{abstract}
	
MIDI--sheet music alignment is the task of finding correspondences between a MIDI representation of a piece and its corresponding sheet music images.  Rather than using optical music recognition to bridge the gap between sheet music and MIDI, we explore an alternative approach: projecting the MIDI data into pixel space and performing alignment in the image domain.  Our method converts the MIDI data into a crude representation of the score that only contains rectangular floating notehead blobs, a process we call bootleg score synthesis.  Furthermore, we project sheet music images into the same bootleg space by applying a deep watershed notehead detector and filling in the bounding boxes around each detected notehead.  Finally, we align the bootleg representations using a simple variant of dynamic time warping.  On a dataset of 68 real scanned piano scores from IMSLP and corresponding MIDI performances, our method achieves a $97.3\%$ accuracy at an error tolerance of one second, outperforming several baseline systems that employ optical music recognition.

\end{abstract}
\section{Introduction}
\label{sec:intro}

This paper tackles the problem of MIDI--sheet music synchronization.  Given a symbolic music representation and its scanned sheet music, the goal is to determine the alignment between each time instant in the symbolic representation and its corresponding pixel location in the sheet music.


Many tools for alignment have been developed in the context of audio synchronization.  The goal of audio synchronization is to find the temporal alignment between two different audio recordings of the same musical piece.  The main technique used to solve this alignment problem is called dynamic time warping (DTW) \cite{DannenbergH03_matching_ICMC}\cite{EwertMG09_HighResAudioSync_ICASSP}\cite{HuDT03_audiomatching_WASPAA}.  DTW consists of four steps: (1) extracting a sequence of features from both audio recordings, (2) computing a cost matrix $C$, where $C_{ij}$ indicates the dissimilarity between the $i^{th}$ frame of recording $1$ and the $j^{th}$ frame of recording $2$, (3) using dynamic programming to calculate a cumulative cost matrix $D$ and backtrace matrix $B$, where $D_{ij}$ indicates the optimal path score from $(0,0)$ to $(i,j)$ given a set of allowable transitions and transition weights, and where $B_{ij}$ indicates the penultimate element in the optimal path, and (4) backtracing through $B$ to determine the lowest cost path through the entire matrix.  Many works have proposed ways to extend or improve upon this basic method, including doing the time warping in an online fashion \cite{dixon2005live}\cite{macrae2010accurate}, estimating the alignment at multiple granularities \cite{MuellerMK06_EfficientMultiscaleApproach_ISMIR}\cite{SalvadorC04_fastDTW}, handling repeats and jumps \cite{FremereyMC10_RepeatsJumps_ISMIR}, handling subsequences or partial alignments \cite{MuellerA08_PathConstrained_ICASSP}\cite{tsai2017make}, dealing with fixed memory constraints \cite{PraetzlichDM16_MsDTW_ICASSP}, and utilizing multiple recordings \cite{ArztWidmer15_RealtimeTrack_ISMIR}\cite{wang2016robust}.


Several previous works have studied the problem of finding correspondences between audio and sheet music.  There are two general approaches to the problem.  The first approach is to use an existing optical music recognition (OMR) system to convert the sheet music into a symbolic (MIDI-like) representation, to collapse the pitch information across octaves to get a chroma representation, and then to compare this representation to chroma features extracted from the audio.  This approach has been applied to synchronizing audio and sheet music \cite{DammFKMC08_MultimodalPresentationofMusic_ICMI}\cite{KurthMFCC07_AutomatedSynchronization_ISMIR}\cite{ThomasFMC12_LinkingSheetMusicAudio_DagstuhlFU}, identifying audio recordings that correspond to a given sheet music representation \cite{FremereyMKC08_AutomaticMapping_ISMIR}, and finding the audio segment corresponding to a fragment of sheet music \cite{FremereyCME09_SheetMusicID_ISMIR}.  A different approach has been explored in recent years: convolutional neural networks (CNNs).  This approach attempts to learn a multimodal CNN that can embed a short segment of sheet music and a short segment of audio into the same feature space, where similarity can be computed directly.  This approach has been explored in the context of online sheet music score following \cite{dorfer2016live}, sheet music retrieval given an audio query \cite{dorfer2016towards}\cite{dorfer2017learning}, and offline alignment of sheet music and audio \cite{dorfer2017learning}.  Dorfer et al. \cite{dorfer2018learning} have also recently shown promising results formulating the score following problem as a reinforcement learning game.

In this paper, we consider the task of MIDI--sheet music synchronization, which can be seen as a variant of the audio--sheet music synchronization scenario. As symbolic (MIDI-like) representations often serve as a bridge between audio and sheet music, MIDI--sheet music synchronization can be regarded as an important intermediate step for more general cross-modal alignment. As mentioned above, traditional approaches typically apply OMR to bridge the modality gap---a step that often introduces severe errors. On the other side, deep learning approaches that try to extract shared feature representations directly from waveforms (audio) and images (sheet music) are promising, but are still in their infancy~\cite{MuellerABDW19_MusicRetrieval_IEEE-SPM}. As the main contribution of this paper, we introduce an approach that avoids an explicit OMR step by working with an explicitly known sparse, binary representation in the image domain. As we will demonstrate, we can convert both sheet image data and MIDI data into this representation using simple logic coupled with a notehead detector. Based on this common binary representation, we show how the alignment problem can then be solved using a simple variant of DTW. In a sense, our approach mimics a deep learning approach, but explicitly introduces a mid-level representation. We hope that our contribution not only sheds a new light into cross-modal alignment, but may also serve as a non-trivial baseline approach for future fully automated procedures.



The paper is organized as follows.  Section 2 describes the proposed algorithm.  Section 3 explains the experimental results.  Section 4 provides an analysis of system performance.  Section 5 concludes the work.

\section{System Description}
\label{sec:system}

There are two inputs to our system: a MIDI file and its corresponding sheet music.  Similar to recent work \cite{dorfer2016towards, dorfer2017learning, dorfer2018learning}, we assume that the sheet music is presented as a sequence of image strips, where each image strip contains a single line of music.  We focus exclusively on piano music in this work, so each line of music consists of a single grand staff containing an upper staff (treble clef) and a lower staff (bass clef).  The image strips may be different sizes, and the staff lines may appear at a different location on each strip.

Our proposed method has three main steps.  The first step is to convert each image strip into a sparse, binary representation in pixel space ($A_i$ in Figure \ref{fig:bootlegSynthBlock}).  We perform this conversion by applying a notehead detector and filling in the predicted bounding boxes around each detected notehead.  This representation is a very crude representation of the score that only contains rectangular floating notehead blobs.  Accordingly, we call this a bootleg representation.  The second step is to project the MIDI data into the same bootleg space ($B_i$ in Figure \ref{fig:bootlegSynthBlock}).  We perform this projection by converting MIDI note onsets into floating notehead blobs that are appropriately placed in pixel space.  The third step is to align the bootleg representations using a variant of DTW (Figure \ref{fig:dtwBlock}).  These three steps are described in detail in the next three subsections.\footnote{Our code and data are available at \url{https://github.com/ttanprasert/sheet-midi-sync}.}

\subsection{Notehead Detection}

\begin{figure}
	\includegraphics[width=\columnwidth]{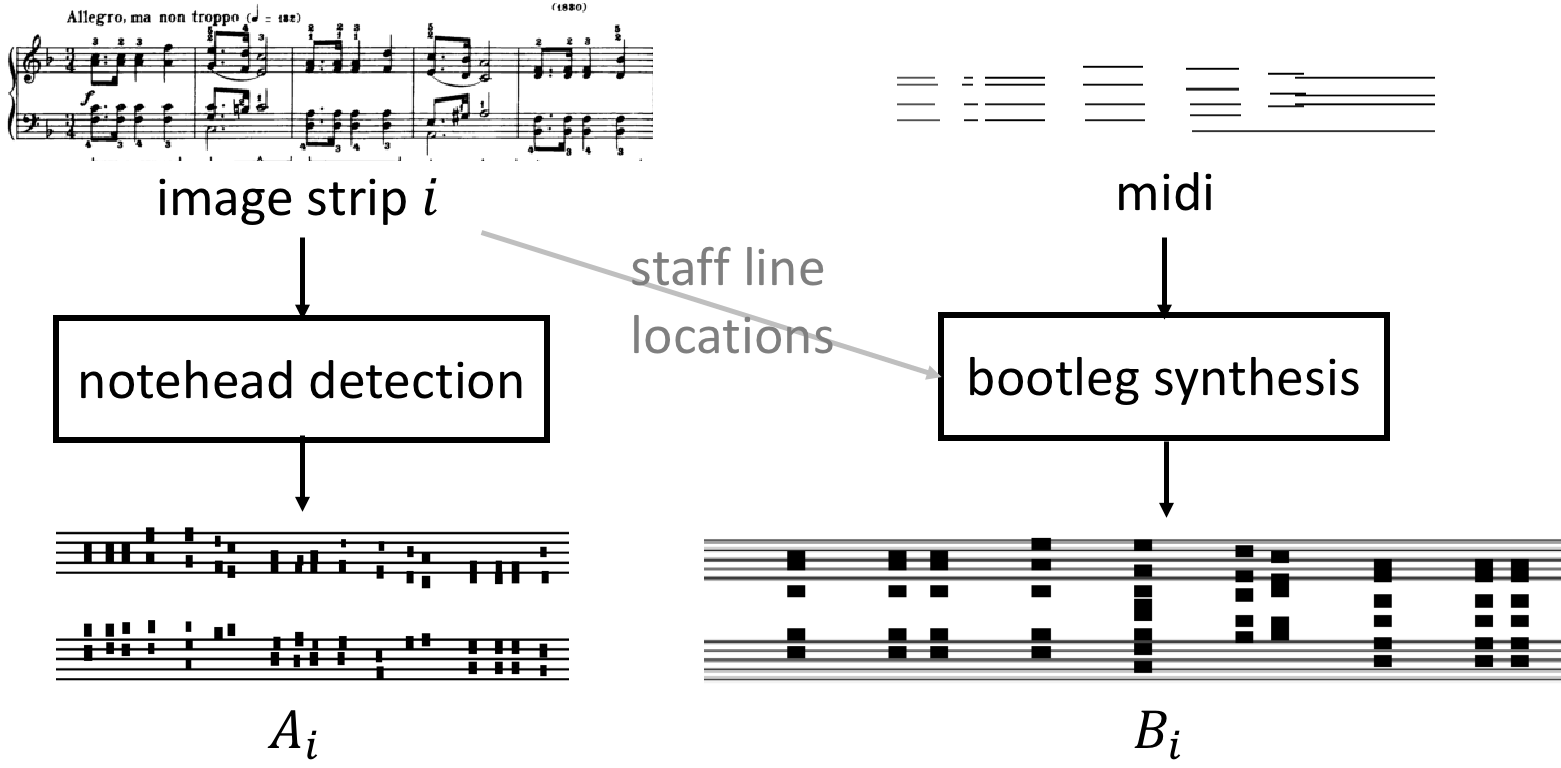}
	\caption{Projecting data to bootleg space.  We convert the image strips and MIDI data into a very crude approximation of the sheet music that only contains floating notehead blobs.  The staff lines in $A_i$ and $B_i$ are shown as a visual aid, but are not included in the bootleg representation.}
	\label{fig:bootlegSynthBlock}
\end{figure}

The first step is to convert each image strip into a bootleg representation.  As shown in Figure \ref{fig:bootlegSynthBlock} (left side), we accomplish this by applying a notehead detector and filling in the predicted bounding boxes around each detected notehead.  The remainder of this subsection describes our notehead detection.

\begin{figure}
	\center
	\includegraphics[width=\columnwidth]{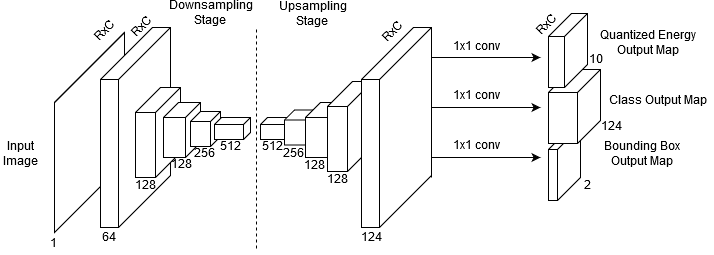}
	\caption{Architecture of the deep watershed notehead detector.  The number below each layer indicates the number of feature maps.  In the downsampling and upsampling stages, the length and width of the feature maps change by a factor of two in each successive layer.  We train on the DeepScores dataset \cite{tuggener2018deepscores} and fine-tune on a small set of manually labeled noteheads in real scanned music.}
	\label{fig:network_architecture}
\end{figure}

Our notehead detector is based on the deep watershed detector recently proposed by Tuggener et al. \cite{tuggener2018deepwatershed} for musical object detection in sheet music.  The deep watershed detector is a fully convolutional network \cite{long2015fully} modified to predict three outputs: (a) a quantized energy output map which indicates the likelihood of having an object at each pixel location, (b) a class output map which predicts which type of object is present at each pixel location (e.g. filled notehead, staff line, treble clef, sharp, quarter rest, etc.), and (c) a bounding box output map which indicates the width and height of an object at that pixel location.  Figure \ref{fig:network_architecture} illustrates the overall architecture of the deep notehead detection network.  The reader is referred to \cite{tuggener2018deepwatershed} for more details.  In \cite{tuggener2018deepscores} and \cite{tuggener2018deepwatershed}, Tuggener et al. show that fully convolutional networks are more suitable for semantic segmentation and detection of tiny objects in sheet music, tasks where (large) object detection methods like Fast R-CNN \cite{girshick2015fast}, Faster R-CNN \cite{ren2015faster}, and YOLO \cite{redmon2016you} fail miserably.

We trained our network on the DeepScores dataset \cite{tuggener2018deepscores}.  This dataset contains approximately 300,000 full pages of synthetically generated musical scores and pixel-level ground truth labels for 124 different symbol classes.  The inputs to the network are 500x500 grayscale image patches that are randomly sampled from the full page images.  The loss function is a linear combination of the losses from the quantized energy output map (cross entropy loss), class output map (cross entropy loss), and bounding box output map (mean squared error).

After training on DeepScores, we fine-tune the network on real scanned sheet music.  For fine-tuning, we manually annotated the location and type of approximately 2200 noteheads in 30 different pages of piano music downloaded from IMSLP.\footnote{\url{https://imslp.org}}  These 30 pages of music were selected to maximize diversity across composers and music publishers, and they are a completely separate set from the data used to evaluate alignment.  Because we only care about detecting noteheads, we disregard all other musical objects in the fine-tuning process.  Because the real scanned music contains a variety of font sizes, we scale each input image to match the staff line spacing in the DeepScores data.  After fine-tuning, the notehead detector achieves a training mean average precision (mAP) of 0.4201 for all notehead types (black notehead, half notehead, and whole notehead).  For reference, in normal object detection tasks (not tiny objects), the state-of-the-art mAP is around $0.4$ to $0.6$.\footnote{\url{http://cocodataset.org/#detection-leaderboard}}  Figure \ref{fig:sampleAlignment} (top half) shows an example of the notehead detector predictions on a section of Brahms Intermezzo Op. 117 No. 2.

\subsection{Bootleg Synthesis}

The second step is to convert the MIDI data into a bootleg representation.  As shown in Figure \ref{fig:bootlegSynthBlock} (right side), we accomplish this by converting note onsets into appropriately placed floating notehead blobs.  This process consists of three key parts.

The first part is determining the staff line coordinate system.  For each image strip, we would like to determine the location of the staff lines in the upper staff and lower staff.  We can accomplish this by computing the row sum of image pixels, convolving the result with comb filters of various sizes (each containing 5 regularly spaced impulses), and identifying the comb filter that yields the strongest response at two non-overlapping staff locations.  This gives us the staff line coordinate system for the upper and lower staves.

The second part is synthesizing and placing noteheads.  Given the coordinate system from an image strip, we convert each MIDI note onset into one or more floating rectangular noteheads.  Note that there is ambiguity when converting from a MIDI note number to a location on a staff.  For example, a MIDI note number of 68 might appear as a G-sharp or an A-flat, which correspond to two different staff locations.  To handle this ambiguity, we can place a larger-than-normal rectangular notehead which overlaps both possible locations.  Furthermore, since notes in the middle register could appear in the right hand or left hand staves, we can simply place two different floating noteheads at both possible locations.  In the visualization of $B_i$ in Figure \ref{fig:bootlegSynthBlock}, for example, you can see that the first chord contains a C4, which produces a notehead in both the upper and lower staves.

\begin{figure}
	\includegraphics[width=\columnwidth]{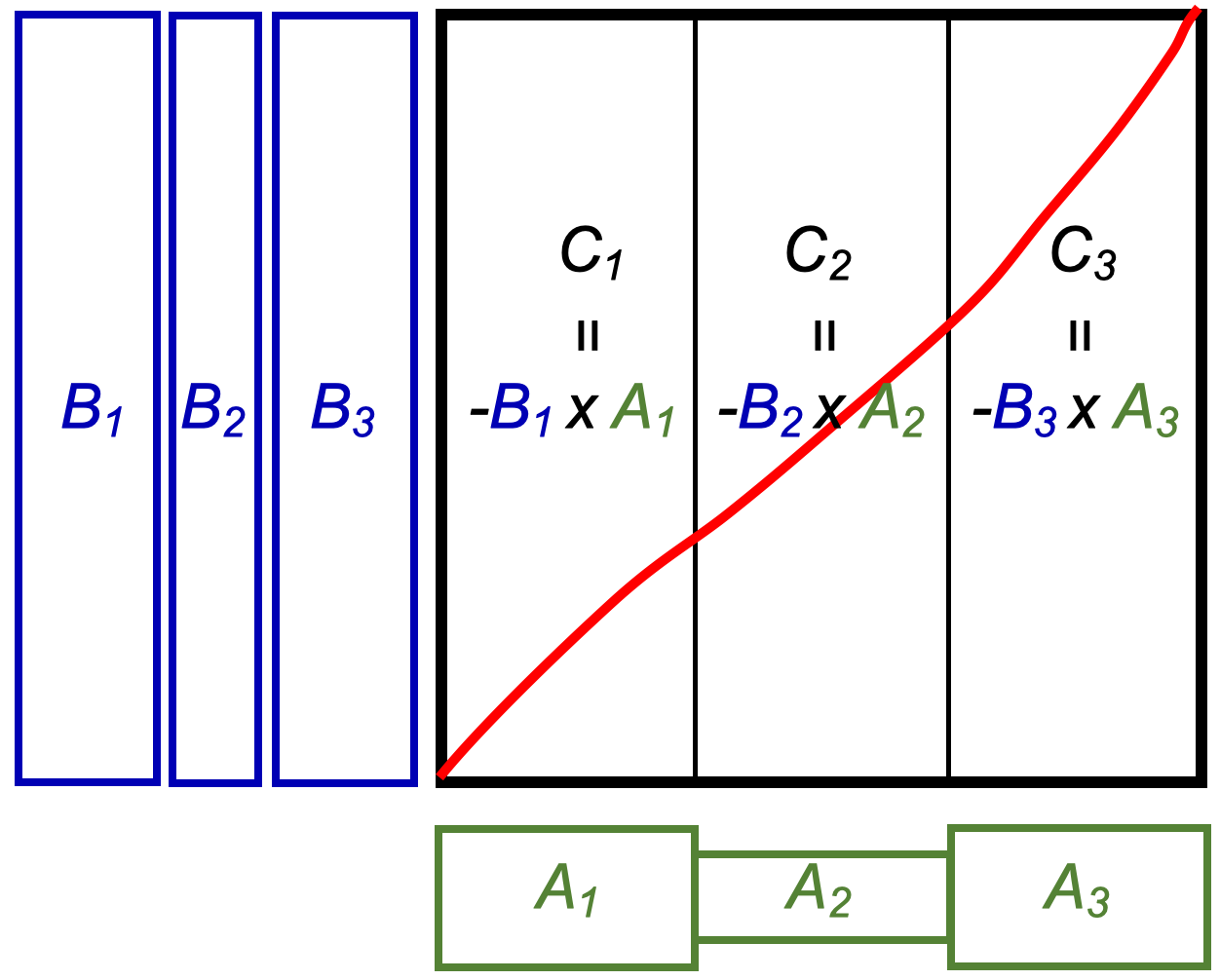}
	\caption{Aligning MIDI-generated bootleg scores with the sheet image-generated bootleg strips.  Each bootleg strip $A_i$ is compared to its corresponding bootleg score $B_i$ to yield a cost matrix block $C_i$.  Note that each $B_i$ is a MIDI-generated bootleg score of the \emph{entire} piece projected onto the staff line coordinate system of strip $A_i$.  The red line indicates an alignment path.}
	\label{fig:dtwBlock}
\end{figure}

The third part is handling timing issues.  One issue is the choice of sampling rate.  When converting the MIDI data into the bootleg representation, we must choose how much time corresponds to a single pixel column.  To avoid extreme time warping, we select this parameter to ensure that the bootleg representation is approximately the same length as the image strips concatenated end-to-end.  Another issue is shortening long pauses.  When there is a fermata in the score, for example, the MIDI performance may slow down in tempo by a factor of three or four.  The sheet music, however, does not reflect this, i.e.,\ the length of the measure in pixels is not elongated by a factor of three or four.  To mitigate this issue, we simply shorten long gaps greater than a fixed threshold to the length of the threshold.  In experiments, we find that system performance is relatively insensitive to this threshold (across an order of magnitude).

Because each sheet image strip $A_i$ has a different size and a different staff line coordinate system, we generate one bootleg score $B_i$ of the \emph{entire} MIDI performance for each image strip $A_i$.  In other words, $B_i$ is the bootleg score representation of the entire MIDI performance projected onto the staff line coordinate system of image strip $A_i$. A schematic illustration of this process is shown in Figure \ref{fig:dtwBlock} for the case of three image strips $A_1$, $A_2$, and $A_3$.  Note that the duration of each $B_i$ ($i\in\{1,2,3\}$) semantically corresponds to the total duration of the concatenation $A_1A_2A_3$, while the pixel height of each $B_i$ matches the height of image strip $A_i$.

\subsection{Block DTW}

The third step is to determine the alignment between the bootleg representations.  Figure \ref{fig:dtwBlock} shows a graphical depiction of this process.  In this example, the sheet music contains three image strips $A_1$, $A_2$, and $A_3$.  The alignment is carried out in two substeps.

The first substep is to calculate the cost matrix ($C_i$) between each bootleg image strip ($A_i$) and its corresponding MIDI-generated bootleg score ($B_i$).  In choosing a suitable cost metric, we must consider the nature of the bootleg representations.  The MIDI-generated bootleg score will have many redundant notes, where (for example) a C4 will appear in both the left and right hand staff systems in order to handle both possibilities.  For this reason, we do not want to penalize the two bootleg representations when they disagree---we only want to reward them when they agree.  One simple cost metric that meets this criteria is a negative inner product, i.e. the $(k,\ell)^\mathrm{th}$ element of $C_i$ indicates ($-1$ times) the number of overlapping black pixels in the $k^\mathrm{th}$ pixel column of $B_i$ and the $\ell^\mathrm{th}$ pixel column of $A_i$.  When black ink shows up in the same vertical pixel position in two pixel columns, it will make the cost more negative.  

The second substep is to perform global DTW.  We assemble the constituent cost matrices $C_i$ into a single global cost matrix (represented as a bold black rectangle in Figure \ref{fig:dtwBlock}).  We then apply DTW with step transitions $\{(1,1), (1,2), (2,1)\}$ and corresponding weights $\{2, 3, 3\}$.  This set of step transitions and weights is a robust, common choice in alignment tasks (e.g. see \cite{muller2015fundamentals}).  The lowest cost path through the global cost matrix is the estimated alignment between the MIDI performance and the sheet music.  The estimated alignment is shown as a red line in Figure \ref{fig:dtwBlock}.

\begin{table}
	\begin{center}
		\scalebox{0.9}{
		\begin{tabular}{|l|c|c|c|c|c|}
			\hline
			Piece & Sh & Meas &  Strips\\
			\hline
			Brahms Fantasia Op117No2 & 4 & 86 & 20,25 \\
			Brahms Fantasia Op116No6 & 3 & 64 & 12,15 \\
			Chopin Mazurka Op30No2 & 6 & 64 & 9,12 \\
			Chopin Mazurka Op63No3 & 6 & 76 & 10,12 \\
			Chopin Mazurka Op68No3 & 6 & 60 & 8,12 \\
			Clementi Sonata Op36No1 mv3 & 2 & 70 & 8,8 \\
			Clementi Sonata Op36No2 mv3 & 2 & 111 & 14,14 \\
			Clementi Sonata Op36No3 mv3 & 2 & 82 & 11,11 \\
			Debussy Children's Corner mv1 & 3 & 76 & 24,25 \\
			Debussy Children's Corner mv3 & 3 & 124 & 23,29 \\
			Debussy Children's Corner mv6 & 3 & 128 & 25,25 \\
			Mendelssohn Op19No2 & 5 & 91 & 12,14 \\
			Mendelssohn Op62No3 & 3 & 48 & 8,10 \\
			Mendelssohn Op62No5 & 3 & 59 & 12,13 \\
			Mozart Sonata No13 mv3 & 4 & 225 & 42,50 \\
			Mozart Sonata No9 mv3 & 3 & 269 & 50,60 \\
			Schubert Impromptu Op90No1 & 2 & 204 & 41,60 \\
			Schubert Impromptu Op90No3 & 2 & 86 & 42,42 \\
			Schubert Op94No2 & 2 & 92 & 17,20 \\
			Tchaikovsky The Seasons - Jan & 2 & 102 & 29,29 \\
			Tchaikovsky The Seasons - Jun & 2 & 99 & 38,40 \\
			Tchaikovsky The Seasons - Aug & 2 & 198 & 24,24 \\
			\hline
		\end{tabular}
	}
	\end{center}
	\caption{Summary of dataset.  For each piece, the table indicates the number of sheet music versions (Sh), the number of measures (Meas), and the minimum \& maximum number of image strips (i.e. lines of music) across the different sheet music versions.}
	\label{tab:datasetSummary}
\end{table}

\section{Experiments}
\label{sec:experiments}

We now summarize our experiments, where we evaluate and compare our bootleg alignment approach with several baseline approaches.  In Section \ref{subsec:setup}, we describe our experimental setup introducing the dataset, the manually generated reference annotations, and the evaluation measure.  In Section \ref{subsec:results}, we then present and discuss our quantitative results.


\subsection{Experimental Setup}
\label{subsec:setup}

The data consists of sheet music scans and MIDI representations for 22 compositions from 8 different composers.  The pieces are all for solo piano, contain no repeats or structural jumps, and span a variety of eras, styles, and lengths.  The sheet music is downloaded from IMSLP and contains digital scans of printed sheet music editions in the public domain.  Note that the choice of using scans of real printed sheet music is a significant departure from other works that focus on synthetically rendered sheet music representations (e.g. \cite{dorfer2018learning, dorfer2017learning}).  In total, there are 68 sheet music scores.  For each composition, we also collected one MIDI performance from online websites.\footnote{\url{www.piano-midi.de} and \url{www.mazurka.org.uk}}  The MIDI performances are symbolic score representations that have been modified to sound like expressive, realistic human performances.  Table \ref{tab:datasetSummary} summarizes the dataset.

The ground truth consists of beat-level annotations.  For the sheet music, we annotate the horizontal pixel location of a subset of beats in each piece, along with the measure number and image strip number.  Because pixel-level annotation of beat locations is very time-consuming, we annotate the beats in $N=40$ measures equally spaced throughout each piece.  For the MIDI performances, we estimate the ground truth beat locations using \url{pretty-midi}\footnote{\url{https://github.com/craffel/pretty-midi}} and manually correct any errors.

To evaluate the system performance, we compare the predicted alignment to the ground truth annotations.  At the ground truth beat locations in the sheet music, we compare the predicted corresponding times in the MIDI to the ground truth timestamps.  Given a fixed error tolerance, we define the error rate to be the percentage of predictions that fall outside of the allowable error tolerance.  By considering a range of error tolerances, we can characterize the tradeoff between error rate and error tolerance.

In total, we have $68$ MIDI-score pairings, and the resulting alignments are evaluated at $10,913$ ground truth beat locations.  Our choice to use scans of real published musical scores has a tradeoff: it places a constraint on the size of the dataset due to the time-consuming nature of annotation, but it is more representative of performance ``in the wild" compared to large synthetic datasets like the Multimodal Sheet Music Dataset (MSMD) \cite{dorfer2018tismir}.  We also evaluate our system on MSMD as a point of comparison.

\begin{figure}
	\includegraphics[width=\columnwidth]{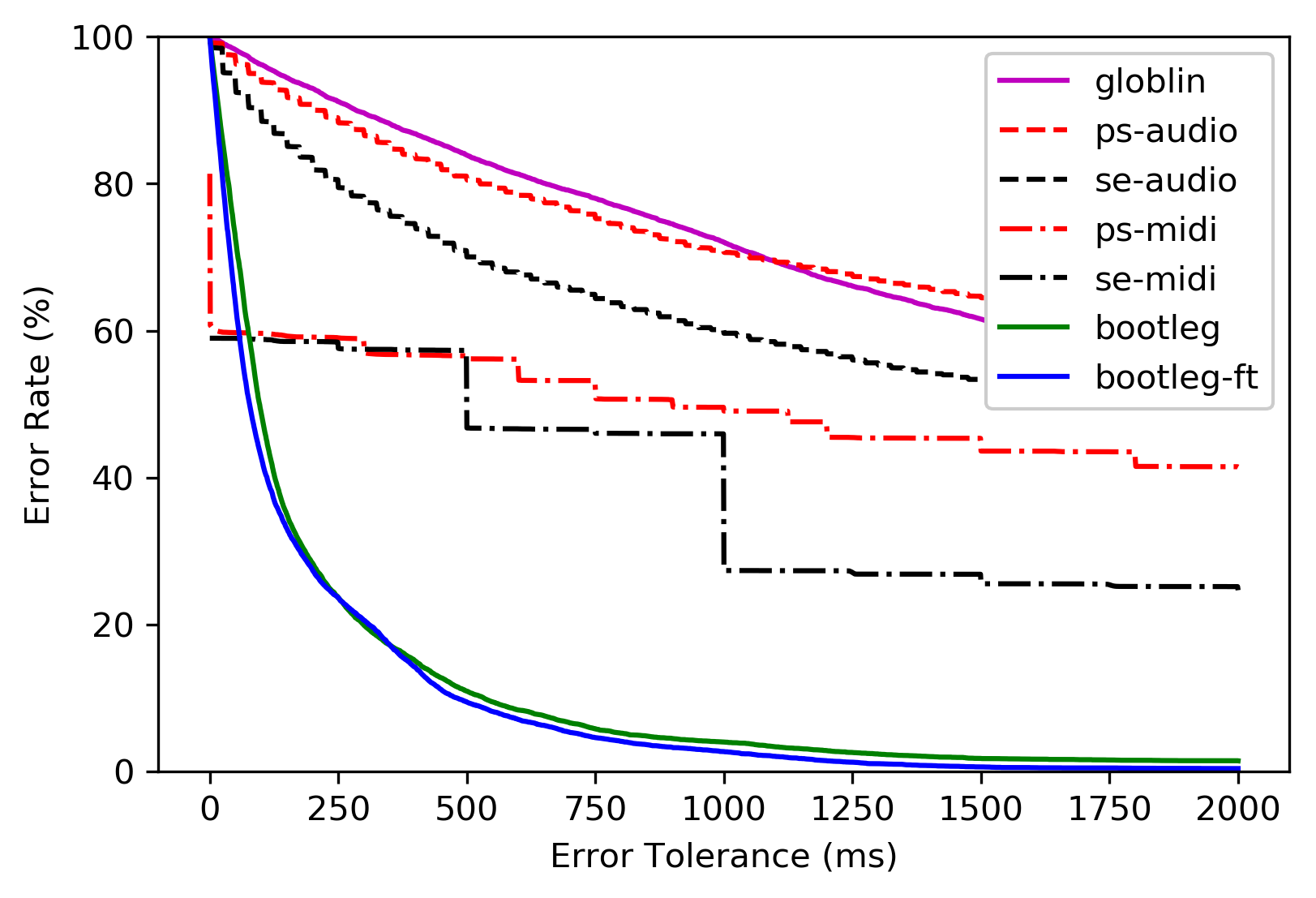}
	\caption{Comparison of baselines to the bootleg system with and without fine-tuning.  The legend lists the systems in order of performance from worst to best.  For a description of the baseline systems, see Section \ref{subsec:results}.}
	\label{fig:errorCurves}
\end{figure}

\subsection{Results}
\label{subsec:results}

We compare our bootleg method (with and without fine-tuning the notehead detector) to five baseline systems.  The first baseline system (`globlin') simply assumes a global linear correspondence between the concatenated sheet image strips and the MIDI performance.  The second and third baseline systems use two different commercial OMR systems (Photoscore\footnote{\url{https://www.neuratron.com/photoscore.htm}} and SharpEye\footnote{\url{http://www.visiv.co.uk}}) to convert the sheet music to MIDI, synthesize the MIDI to audio, and then perform audio--audio alignment using DTW on chroma features with $25$ ms hop size.\footnote{We also experimented with doing DTW directly on a piano roll representation of the MIDI data, but found that the results were always worse than synthesizing to audio and aligning chroma features.}  These baselines are abbreviated as `ps-audio' and `se-audio' in Figure \ref{fig:errorCurves}.  The fourth and fifth baseline systems use the same two OMR systems to convert the sheet music to MIDI, estimate the beat locations using \url{pretty-midi}, and then assume a 1-to-1 correspondence between beat locations in both MIDI files.  Note that the OMR system can have many recognition errors but still have perfect alignment if it can simply interpret barlines and beats correctly.  These two baselines are abbreviated as `ps-midi' and `se-midi' in Figure \ref{fig:errorCurves}.\footnote{Note that the stairstep shape of the `se-midi' system comes from the fact that SharpEye always renders its OMR-generated MIDI at 120 BPM, so that missing or extra beats correspond to errors at integer multiples of 500 ms.}  


There is one important issue to mention about evaluating the OMR baseline systems.  Because PhotoScore and SharpEye do not retain the connection between sheet music pixel location and corresponding MIDI time, there is no reliable way to automatically infer ground truth beat locations in the OMR-generated MIDI.  Thus, it was necessary to manually annotate the ground truth beat locations in the OMR-generated MIDI on every sheet music score, so that the predicted alignment can be evaluated.  This is a very time-consuming process, and clearly not sustainable for large-scale evaluations.  However, the benefit of these results is a fair comparison to commercial OMR systems over a reasonably diverse data set.

\begin{figure}
	\includegraphics[width=\columnwidth]{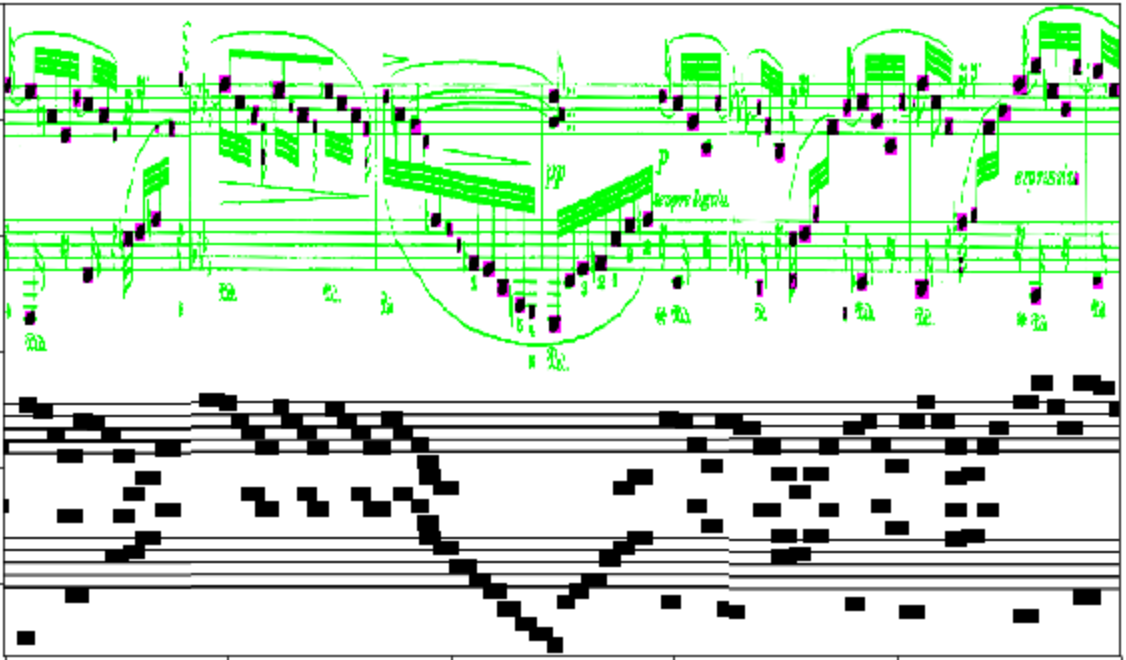}
	\caption{An example of the predicted alignment produced by the bootleg system.  The upper half of the figure shows the original score with the detected noteheads overlaid.  The bottom half of the figure shows the aligned MIDI-generated bootleg score.  This figure is best visualized in color.}
	\label{fig:sampleAlignment}
\end{figure}

Figure \ref{fig:errorCurves} compares the performance of our proposed bootleg approach (with and without fine-tuning) against the five baseline systems.  There are three things to notice about Figure \ref{fig:errorCurves}.


First, the bootleg method (`bootleg') outperforms the baselines by a wide margin.  For example, at 500 ms error tolerance the bootleg systems achieve error rates around $10\%$, whereas the best performing baseline system achieves an error rate of $47\%$.  Similarly, at 1000 ms error tolerance the bootleg systems achieve $3-4\%$ error rate, whereas the best baseline system has a $27\%$ error rate.

Second, fine-tuning the notehead detector on sheet music scans (`bootleg-ft') shows demonstrable improvement.  For example, at 1000 ms error tolerance the fine-tuning improves the error rate from $4.0\%$ to $2.7\%$, and at 100 ms error tolerance the fine-tuning improves the error rate from $48.8\%$ to $42.8\%$.  We already know that fine-tuning will always improve results.  The key observation here is that we can significantly improve results even with an extremely small dataset (2200 noteheads).

Third, the bootleg systems achieve very low asymptotic error rates.  Whereas the best-performing baseline system achieves an error rate of $24.7\%$ at a 2000 ms error tolerance, the fine-tuned bootleg system achieves a $0.4\%$ error rate.  So, the bootleg alignments are reliable, at least on the data in our experiments.


\section{Further Analysis}

In this section, we further investigate the proposed system through three different types of analyses.

\subsection{Visualization of Alignments}

The first method of analysis is to create a visualization that shows the predicted alignment between the bootleg representations.  Figure \ref{fig:sampleAlignment} shows an example from a section of Brahms Intermezzo Op. 117 No. 2.  In the upper half, the detected notehead regions are overlaid on top of the original score for ease of visualization.  The bottom half contains the aligned MIDI-generated bootleg score.  We can see that most noteheads are correctly detected, and the two bootleg representations match well.

By looking at example visualizations, we discovered two weaknesses in our system.  The first weakness is that the notehead detector performs poorly on half noteheads and whole noteheads.  This is due to the fact that the training data was highly imbalanced towards filled noteheads.  The second weakness is that the staff line detection occasionally fails, which leads to poor alignments on the entire strip.  Interestingly, the system is fairly robust to clef changes, mainly because clef changes usually only occur in one staff but not in both staffs simultaneously.

\begin{figure}
	\includegraphics[width=\columnwidth]{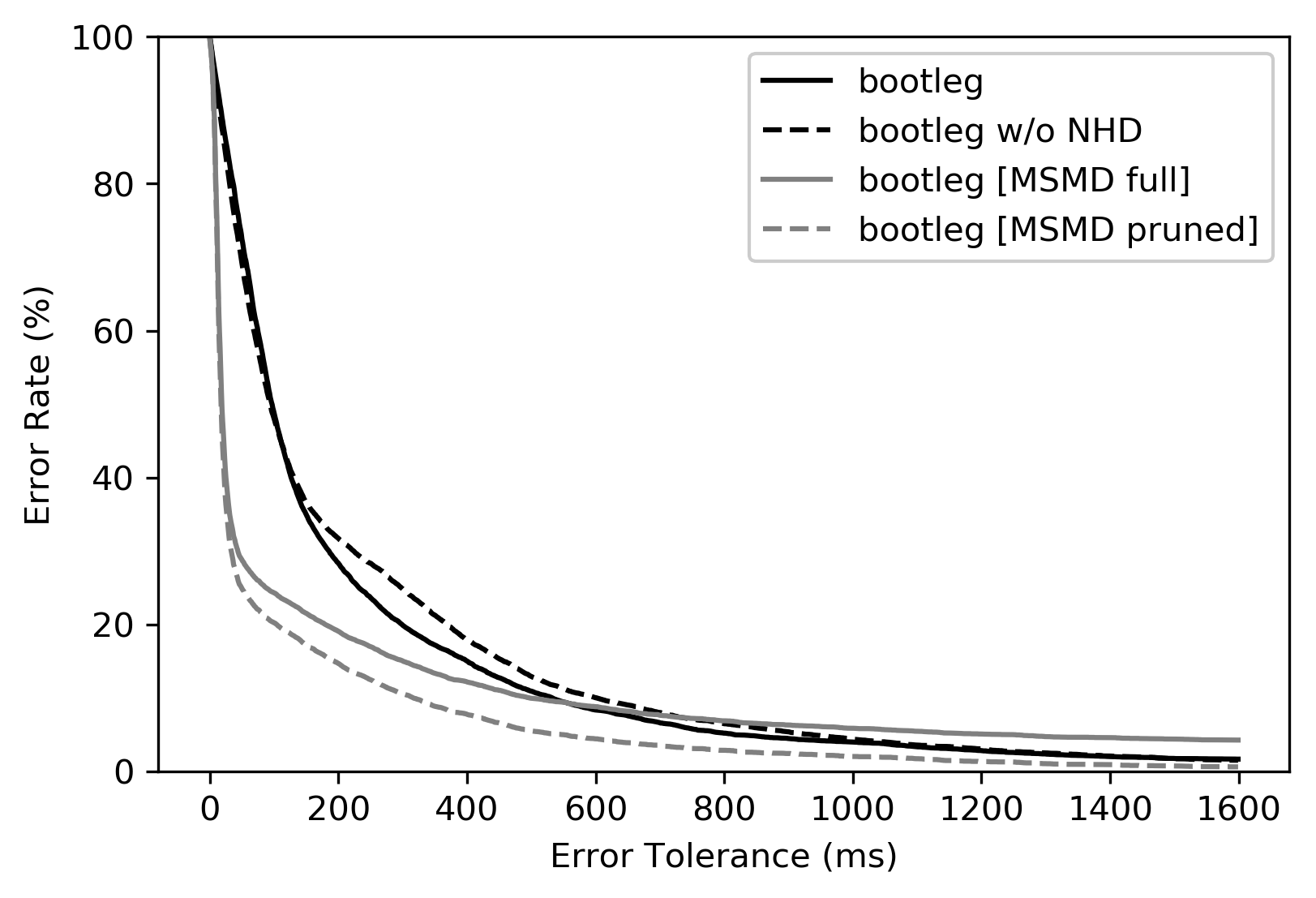}
	\caption{This figure shows two comparisons:  (a) The performance of the bootleg system on real scanned sheet music vs. synthetically generated sheet music (see Section \ref{subsec:synthVsReal}).  (b) The performance of the bootleg system with and without notehead detection (see Section \ref{subsec:importanceNHD}).}
	\label{fig:errorCurvesAnalysis}
\end{figure}

\subsection{Synthetic vs. Real Data}
\label{subsec:synthVsReal}

The second analysis investigates the question: ``How well does the proposed system work on synthetic data vs. real data?"  As mentioned before, we chose to evaluate the baseline systems on real data (i.e. scans of music from IMSLP) rather than synthetically rendered data so that we can assess performance ``in the wild."  The primary drawback of using real data is that it needs to be manually annotated, which is very costly and limits the practical size of the evaluation dataset.  To see how well our system performs on synthetic data, we also ran a large-scale evaluation on the Multimodal Sheet Music Dataset (MSMD) \cite{dorfer2018tismir}.  

Figure \ref{fig:errorCurvesAnalysis} compares the performance of the bootleg system on the real dataset (black solid line) and the test set from MSMD (gray solid line).  We can see that the synthetic data is ``easier" to align, especially at lower error tolerances.  However, the performance on MSMD leveled off at a much higher error rate for large error tolerances (e.g. $4.4\%$ error at 1500 ms tolerance vs. $1.8\%$ error on the real dataset).  Upon further investigation, we found that these errors came from a small set of six pieces that all had one of two peculiar characteristics: (a) they consisted of almost all half or whole notes throughout the entire piece (e.g. Erik Satie's \textit{Gymnopedies}), or (b) they had very frequent time signature changes (on average every $1.4$ measures for the two relevant pieces).  The first characteristic will cause the bootleg system to fail because of the imbalance in notehead detection training data, and the second will cause extreme time warping in the alignment stage.  While these pieces might be considered extreme or unusual in this regard, they nonetheless provide additional insight into failure modes of the bootleg approach.  When we removed this set of six pieces from evaluation, the error curve falls significantly (dotted gray line in Figure \ref{fig:errorCurvesAnalysis}).  We can interpret the gap between the dotted gray curve and the solid black curve as the performance loss when transitioning from synthetic data to real scanned sheet music.

\subsection{Importance of Notehead Detection}
\label{subsec:importanceNHD}

The third analysis investigates the question: ``How much does system performance depend on notehead detection?"  To answer this question, we simply removed the notehead detection from our system and directly aligned the MIDI-generated bootleg scores to the raw image strips.  We expect the performance to be worse without notehead detection because the raw sheet music will contain many symbols that the MIDI-generated bootleg score does not have: note stems, rests, accidentals, etc.  These additional symbols introduce noise that can lead to poor alignments.

Figure \ref{fig:errorCurvesAnalysis} compares the performance of the bootleg system with (solid black line) and without (dotted black line) notehead detection on the real dataset.  Surprisingly, the system without notehead detection performs only marginally worse, and it approaches approximately the same error rate at high error tolerances.  This suggests a way to significantly reduce the complexity of the system without sacrificing much performance.

\section{Conclusion}

We investigate the MIDI--sheet music synchronization problem as an important intermediate step for cross-modal alignment.  Because OMR is a difficult task and may not be needed for music alignment, we avoid the need for OMR by introducing a mid-level representation called a bootleg score.  We project the MIDI data into bootleg space using the rules of Western musical notation, and we project the sheet music into bootleg space by applying a deep watershed notehead detector.  Once the MIDI and sheet music have been projected to this bootleg representation, the alignment can be performed using a simple variant of DTW.  We evaluate the proposed system on scans of real published piano scores.  Our results indicate that the proposed approach works well for piano music, and it outperforms several baseline systems based on optical music recognition.  Our approach may serve as a non-trivial baseline approach for future end-to-end learning approaches.

\section{Acknowledgments}
Thitaree Tanprasert and Teerapat Jenrungrot were supported by the Class of 1989 Summer Experiential Learning Fund, the Vandiver Summer Experiential Learning Fund, and the Norman F. Sprague III, M.D. Experiential Learning Fund established by the Jean Perkins Foundation.  We gratefully acknowledge the support of NVIDIA Corporation with the donation of the Tesla K40 GPU used for this research.  Meinard M\"uller was supported by the German Research Foundation (DFG MU
2686/12-1). The International Audio Laboratories Erlangen are a joint
institution of the Friedrich-Alexander-Universit\"at Erlangen-N\"urnberg
(FAU) and Fraunhofer Institut f{\" u}r Integrierte Schaltungen IIS.

\bibliography{sheetMidiSync}

%
%

\end{document}